\documentstyle[12pt,aasms]{article}
\def\ecs{ergs cm$^{-2}$ sec$^{-1}$}
\def\es{ergs sec$^{-1}$ }

\def\lae{\mathrel{<\kern-1.0em\lower0.9ex\hbox{$\sim$}}}
\def\gae{\mathrel{>\kern-1.0em\lower0.9ex\hbox{$\sim$}}}

\received{: August 4, 1998}
\accepted{: October 12, 1998}

\slugcomment{Astrophysical Journal, in press}

\begin{document}

\title{ASCA Observations of the Starburst-Driven Superwind Galaxy NGC 2146: 
Broad Band (0.6 - 9 keV) Spectral Properties}


\author{Roberto Della Ceca\altaffilmark{1}, 
        Richard E. Griffiths\altaffilmark{2},
        Timothy M. Heckman\altaffilmark{3},
        Matthew D. Lehnert\altaffilmark{4} and
        Kimberly A. Weaver\altaffilmark{5}}


\altaffiltext{1}{Osservatorio Astronomico di Brera, 
     20121 Milano, Italy. E-mail: rdc@brera.mi.astro.it} 
\altaffiltext{2}{Physics Department, Carnegie Mellon University, 
     Pittsburgh, PA 15213-3890, USA. E-mail: griffith@astro.phys.cmu.edu} 
\altaffiltext{3}{Department of Physics and Astronomy, The Johns Hopkins 
     University, Baltimore, MD 21218, USA. E-mail: heckman@eta.pha.jhu.edu}
\altaffiltext{4}{Sterrewacht Leiden, Postbus 9513, 2300 RA Leiden, 
     The Netherlands. E-mail: lehnert@strw.leidenuniv.nl}
\altaffiltext{5}{NASA/Goddard Space Flight Center, Code 662, Greenbelt, 
      Maryland 20771, USA. E-mail: kweaver@milkway.gsfc.nasa.gov}



\begin{abstract}

We report ASCA GIS and SIS observations of the nearby (D = 11.6 Mpc),
nearly edge-on, starburst galaxy NGC 2146.  These X-ray spectral data
complement ROSAT PSPC and HRI imaging discussed by Armus et al., 1995.

The broad band (0.6-9 keV) X-ray spectrum of NGC 2146 is best described
by a two component model: the soft X-ray emission with a Raymond-Smith
thermal plasma model having a temperature of kT $\sim 0.8$ keV; the
hard X-ray emission with a thermal plasma model having kT $\sim 8$ keV
or a power-law model having a photon index of $\sim 1.7$. We do not
find compelling evidence of substantial excess absorption above the
Galactic value.  The total luminosities of NGC 2146 in the soft (0.5 -
2.0 keV), hard (2-10 keV) and broad (0.5-10.0 keV) energy bands are
$\sim 1.3 \times 10^{40}$, $\sim 1.8 \times 10^{40}$ and $\sim 3.1
\times 10^{40}$ \es, respectively.  The soft (hard) thermal component
provides about 30\% (70\%) of the total luminosity in the 0.5 - 2.0 keV
energy band, while in the 2-10 keV energy range only the hard component
plays a major role.

The spectral results allow us to set tighter constraints on the
starburst-driven superwind model, which we show can satisfactorily
account for the luminosity, mass, and energy content represented by the
soft X-ray spectral component.  We estimate that the mass outflow rate
($\sim$ 9 M$_{\odot}$ per year) is about an order of magnitude greater
than the predicted rate at which supernovae and stellar winds return
mass into the interstellar medium and, therefore, argue that the flow
is strongly ``mass-loaded" with material in and around the starburst.
The estimated outflow velocity of the hot gas is close to the escape
velocity from the galaxy, so the fate of the gas is not clear.  We
suggest that the hard X-ray spectral component is due to the combined
emission of X-ray binaries and/or young supernovae remnants associated
with the starburst.

\end{abstract}


%
%

\section{Introduction}

In the past few years there has been increasing theoretical and
observational evidence demonstrating that the collective effect of
multiple supernovae and stellar winds in starburst galaxies can drive a
galactic-scale outflow of gas, hereafter a ``superwind".  Such
phenomena have been hypothesized to be the mechanism whereby
proto-elliptical galaxies and bulges are cleared of their nascent
interstellar medium, to be responsible for establishing the
metallicity-radius relation or the mass-metallicity relation among
galaxies, and for chemically-enriching and heating the inter-galactic
medium (see Heckman, Lehnert and Armus, 1993; Bland-Hawthorn 1995 for
recent reviews).  The study of this process is therefore of fundamental
importance for understanding many areas of current observational
cosmology.  Optical and radio observations can provide only indirect
evidence of superwind.  On the contrary, X-rays directly probe the
physical condition of the hot wind material. Such data thus play a key
role in the detection, study and investigation of the process of
starburst-driven outflows.

NGC 2146 is a nearby (D=11.6 Mpc, at this distance $1^{\prime} \simeq
3.37$ kpc) SB(s)ab, nearly edge-on ($i \sim 63^{o}$), starburst galaxy
with multi-wavelength evidence for driving a ``superwind".  The
absolute magnitude of NGC 2146 is $M_B = -19.2$ (B$_T$ = 11.38; $L_B
\simeq 7.5 \times 10^9 L_{B,\odot}$); the optical size of the galaxy
(at the limiting surface brightness of 25 B magnitude per square
arcsec, after correction for Galactic extinction) is about $6.0 \times
3.4$ arcmin ($\simeq \ 20.2\times 11.5$ kpc); the position angle of the
major axis is about 56$^o$ (de Vaucouleurs et al., 1991).  Its large
$H_{\alpha}$ luminosity ($L_{H_{\alpha}} \simeq 1.5\times 10^{41}$ \es)
and strong, warm infrared emission ($L_{(1-1000 {\mu}m)} \simeq 1.5
\times 10^{44}$ \es; $S_{(60 {\mu}m)} \sim 0.76 S_{(100 {\mu}m)}$) are
similar to those observed in other  starburst galaxies with suspected
``superwind" activity (see Lehnert and Heckman, 1995; 1996 and
references therein).  Armus et al., 1995 (hereafter A95) discuss in
detail the ``superwind" hypothesis by using optical broad and narrow
$H_{\alpha}$ images, long-slit optical spectra, 6 cm radio data and
X-ray (ROSAT PSPC and ROSAT HRI) data.  We summarize below the X-ray
properties of this galaxy as derived from the ROSAT PSPC and HRI data.

The ROSAT PSPC image  ($E \simeq 0.2-2.4$ keV) show a large X-ray
nebula with a half-light diameter of $\sim 1^{\prime}$ ($\sim 3.37$
kpc), a maximum size of $\sim 4^{\prime}$ ($\sim 13.5$ kpc) and a total
unabsorbed flux (luminosity) of $f_{x_{0.2-2.4 keV}} = 1.1 \times
10^{-12}$ \ecs ($L_{x_{0.2-2.4 keV}} \simeq 1.8 \times 10^{40}$ \es).
This nebula, that extends well beyond the starburst in the NE-SW
direction (approximately along the minor axis of NGC 2146), is much
larger than the starburst ridge seen at 6 cm and seems to be associated
with a region of $H_{\alpha}$ and dust filaments seen in optical
images; it probably traces the expanding superwind.  The inner ($<
0.5^{\prime}$) X-ray nebula is resolved by the ROSAT HRI into at least four
point sources, a nuclear ``spur'', and strong diffuse emission over the
central arcminute.  The X-ray ``spur'' follows the same position angle
along the optical major axis as the 6cm radio ridge. The four resolved
point sources  account for 30\% -- 50\% of the total flux in the HRI.
The ROSAT PSPC X-ray spectrum is best described by a soft thermal
plasma (kT $\sim 0.4 - 0.5$ keV) plus a ``high energy'' component
consistent with either a power law (photon index =1.7) or a thermal
bremsstrahlung (kT $>$ 5 keV), which dominates above 1 keV.  Hovewer,
the limited energy range and spectral resolution of the ROSAT PSPC
prevents a detailed investigation of the X-ray spectral properties of
NGC 2146.

For this reason we observed NGC 2146 with ASCA with the specific aims of: 
a) verifying spectroscopically the presence of multiple components in the 
   X-ray emission;
b) measuring their respective luminosities;
c) evaluating their proportional contribution in the soft ($E<2$ keV) and 
   hard ($E>2$ keV) energy bands and, therefore, 
d) setting tighter constraints on the starburst-driven superwind model.
   
This paper is organized as follows.  In section 2 we present the ASCA
data; their spectral analysis is reported in section 3. Section 4
contains a discussion of the results.  Finally, a summary of our
conclusions is presented in section 5.


\section{ASCA Observations and data preparation}

NGC 2146 was observed on March 26, 1997 by the ASCA satellite (Tanaka,
Inoue and Holt, 1994).  The spectral data were extracted using version
1.3 of the XSELECT software package and version 4.0 of FTOOLS (supplied
by the HEASARC at the Goddard Space Flight Center).  Good time
intervals were selected by applying the ``Standard REV2 Screening"
criteria as reported in chapter 5 of the ASCA Data Reduction Guide (rev
2.0).  For the GIS instruments we have combined HIGH, MEDIUM and LOW
bit rate data; for the SIS instruments we have combined HIGH and MEDIUM
bit rate data.   The SIS was operated in 1 CCD `Bright' and `Faint'
modes during the observation.  Faint mode data were converted to Bright
mode and Hot and Flickering pixels were removed. From an inspection of
the light curves, we also rejected times of high count rate due to
background variations.  The total effective exposure time was 60938 sec
for SIS0, 60817 sec for SIS1 and 66417 sec for GIS2 and GIS3.

Source counts were extracted from a circular region around the centroid
of X-ray emission from NGC 2146.  This region has a 3.6 arcmin radius
for SIS0 and SIS1 (the maximum allowed due to the position of NGC 2146
within the CCD chips) and a 6 arcmin radius for GIS2 and GIS3.  For the
SIS, background counts were taken from source-free rectangular regions
within the CCD chip that contained NGC 2146.  For the GIS, background
counts were taken from source-free circular regions close to NGC 2146.
Source and background counts were extracted in the ``Pulse Invariant''
(PI) energy channels, which have been corrected for spatial and
temporal variations of the detector gain.  We also checked several
regions for differences in the background counting rate, but found no
significant spatial variations.  The total net counts (hereafter ``net
counts" refer to the background subtracted counts) from NGC 2146 are
$\sim 2150$ for SIS0, $\sim 1900$ for SIS1, $\sim 1700$ for GIS2 and
$\sim 2100$ for GIS3.  The net counts represent about $67\%$
(SIS0,SIS1) and $58\%$ (GIS2,GIS3) of the total gross counts in the
source region.

We exclude from the spectral analysis all data with E $>$ 9 keV and E
$<$ 0.6 keV.  The upper limit is chosen because background dominates at
these energies; the lower limit is chosen because the GIS data do
not provide sufficient statistics and because the SIS calibration is
not well known below this energy due to an ill-defined modeling of the
O K edge in the response matrix (see Dotani et al., 1996) and
uncertainties in the corrections for the loss of photons due to charge
transfer.  For the GIS data we use the detector Redistribution Matrix
Files (RMF) gis2v4$\_$0.rmf and gis3v4$\_$0.rmf, respectively.  For the
SIS data the RMF files were obtained using the FTOOLS task SISRMG.  The
Ancillary Response Files (ARF) were created with the version 2.72 of
the FTOOLS task ASCAARF  at the location of NGC 2146 in the detectors
(see George et al., 1992 for the definition of the RMF and ARF
calibration files).

In order to improve the statistics we produce a combined SIS spectrum
(SIS 0 and SIS 1 were added, S01 hereafter) and a combined GIS spectrum
(G23 hereafter).  The combined spectra and their respective background
and response matrix files have been obtained following  the recipe
given in the ASCA Data Reduction (rev 2.0) Guide (see sections 8.9.2
and 8.9.3 and reference therein).  In order to use $\chi^2$ statistic
in the spectral fitting procedure (see Section 3), S01 and G23 were
rebinned to give at least 30 total counts (source + background) per
energy bin.

In section 3 we also use the ROSAT PSPC data in order to test for
intrinsic (i.e. inside NGC 2146) soft X-ray absorption.  NGC 2146 was
observed with the PSPC  in pointed mode (with the object at the center
of the detector) on 1992 February 25 for a total exposure time of 5799
sec.  The data were retrieved from the ROSAT archive (ROR=700440).
Source counts were extracted from a circular region of $\sim 2.5$
arcmin radius around the centroid of X-ray emission from NGC 2146.
Background counts were extracted from an annular region of inner radius
$\sim 6$ arcmin and outer radius $\sim 8$  arcmin around NGC 2146.  A
total of about 400 net counts were accumulated from the source.  All
data with an energy above 0.2 keV were included in the spectral
analysis; the PSPC spectrum was rebinned to give at least 15 net counts
per bin.  It is worth noting that these data are the same as used by
A95.

Spectral analysis (see Section 3) has been performed using version 9.0
of the XSPEC software package (Arnaud, 1996).  The SIS and GIS data are
fitted jointly and the model normalization(s) for each data set are
allowed to be an independent parameter in order to take into account
differences in the absolute calibration of the instruments.


\section{Spectral Analysis}

X-ray observations of the brightest starburst galaxies (e.g.  NGC 253
and M82, Moran and Lehnert, 1997; Ptak et al., 1997; Dahlem, Weaver and
Heckman 1998; Persic et al., 1998; NGC 3310 and NGC 3690, Zezas,
Georgantopoulos and Ward, 1998) as well as of low-mass star-forming
systems (e.g. NGC 1569 and NGC 4449; Della Ceca et al., 1996; Della
Ceca, Griffiths and Heckman, 1997) have shown that the X-ray spectra of
this class of objects are very complex, displaying at least two or
three spectral components. This is not surprising given that X-rays
from starburst galaxies are expected to derive from different classes
of sources like stars, accreting binary star systems, supernova
remnants, diffuse hot gas and outflowing winds, and inverse Compton
emission.

Single-component models do not provide an adequate description of NGC
2146.  A single-temperature thermal model (Raymond-Smith or Mekal) and
a single power law model, both modified by Galactic absorption
($N_{H_{Gal}}= 7.3\times 10^{20}$ cm$^{-2}$, see A95) are rejected at a
high confidence level ($\geq 99.99\%$).  For the thermal models, the
abundance of the X-ray emitting gas is fixed at the solar value.
Allowing $N_H$ and/or the abundance of the X-ray emitting gas to vary
as free parameters (with the condition $N_{H_{fit}} > N_{H_{Gal}}$)
does not produce a better fit.
From the visual inspection of the residuals to the best fit
Raymond-Smith model for each detector (see Figure 1) we note that the
largest discrepancies are not concentrated around expected emission
lines, as in the case of the high signal-to-noise SIS spectra of M82
reported and discussed in Moran and Lehnert (1997). Rather,
discrepancies occur in the shape of the spectrum for E$\lae 2$ keV,
which is a clear indication that a more complex spectrum is needed to
describe the X-ray spectral properties of NGC 2146.  Similar results
are obtained for the power-law model or for the Mekal model.

We then tried the following two-component spectral models:  1) a
two-temperature Raymond-Smith model (hereafter the Ray+Ray model) and
2) a Raymond-Smith model plus a power law model (hereafter the Ray+Po
model).  These trial models were filtered by Galactic absorption
($N_{H_{Gal}} = 7.3\times 10^{20}$ cm$^{-2}$); the abundance of
the X-ray emitting gas was fixed at the solar value.  The results are
reported in Table 1; the folded spectra and the residuals to the best
fit Ray+Po model for each detector are displayed in Figures 2 (G23) and
3 (S01). Similar results are obtained for the Ray+Ray model.  As
evident from the residuals shown in Figures 2 and 3 and from the
$\chi^2_{\nu}$ in Table 1, both models are a good representation of the
ASCA data.  The spectral parameters derived from single-instrument fits
are in good agreement, within the respective errors, with those
obtained from the joint SIS+GIS fits.
We also obtain similar results if we replace the Raymond-Smith thermal
model with the Mekal thermal model, showing that the fits are
insensitive to the detailed physics of the model.

In order to test for the presence of intrinsic absorption, we include
the ROSAT PSPC data in the spectral analysis; the inclusion of these
data allows us to extend the coverage of the NGC 2146 X-ray spectra down
to E$\sim 0.2-0.3$ keV.  The combined ASCA+ROSAT PSPC data set are
fitted with the two-component models discussed above but the absorbing
column density is left as a free parameter.  The results are reported
in Table 1.  The best fit spectral parameters are in good agreement
with those derived from ASCA alone; furthermore the absorbing column
density along the line of sight is not significantly in excess above
the Galactic value.

Moran and Lehnert (1997) by studying the high quality ASCA + ROSAT PSPC data 
of the prototype starburst galaxy M82 have found that the 
(0.1 - 10 keV) spectrum of this galaxy is best described by a three 
components model composed of:
a) a very soft Raymond-Smith thermal component with kT $\simeq$ 0.3 keV;
b) an absorbed Raymond-Smith thermal component with kT $\simeq$ 0.6 keV and
c) a heavily absorbed ($N_H \sim 10^{22}$ cm$^{-2}$)
   hard component that can be described either by a power-law with photon 
   index equal to 1.7 or by a thermal bremsstrahlung with kT $\simeq$ 18 keV.
The observed 0.1-10 keV flux from M82 is dominated from this latter component.
We have tried to fit the ASCA + ROSAT PSPC data of NGC 2146 using a
different absorbing column density for the hard component.  The best
fit spectral parameters for the original two components model(s) are in
very good agreement with those reported in Table 1, the absorbing
column density of the hard component is consistent with zero and the
$\chi ^{2}$ is not improved with the inclusion of this additional
parameter.  So, with the current statistic we are unable to determine the
possible excess absorption of the hard component, as found by Moran and
Lehnert (1997) for M82.

Finally, we have also tried to fit the ASCA + ROSAT PSPC data  with the
abundance as free parameter for the soft component; unfortunately our
data are not able to set any useful constraint on the abundance.

To summarize, the broad band X-ray spectrum of NGC 2146
reveals the presence of at least two spectral components: the soft
component is best described by a Raymond-Smith thermal model with a
temperature of $\sim 0.8$ keV, while the hard component can be
described by a thermal model with a temperature of $\sim 8$ keV or by a
power-law model with photon index of $\sim 1.7$.  There is no
compelling evidence of intrinsic absorption inside NGC 2146.
The unfolded spectrum of the ASCA + ROSAT PSPC data set (Ray+Po spectral model) is shown in Figure 4.

We stress that the two-component spectral models reported above must be
considered as a ``first order'' approximation of the real spectral
energy distribution in NGC 2146. As already discussed at the beginning
of section 3, X-ray emission from starburst galaxies is expected to
derive from several processes and several locations; therefore, more
complex X-ray spectra should be considered more realistic. These are
the limitations of the current data sets.  Spatially resolved
spectroscopy will be possible with the instrumentation on board AXAF,
JET-X and XMM, allowing us to investigate in deeper detail the
spectral properties of NGC 2146 and similar objects.


\section{Discussion}

\subsection{Fluxes and Luminosities}

We report in Table 2 the unabsorbed X-ray fluxes and luminosities of
the spectral components in the soft (0.5 - 2.0 keV) and hard (2.0 -
10.0 keV) energy range for the two spectral models discussed above. For
the normalization of the spectral components we have used a mean value
between the G23 and the S01 normalizations (see Table 1).  The total
unabsorbed luminosity of NGC 2146 in the soft, hard and broad (0.5-10.0
keV) energy band is $\sim 1.3 \times 10^{40}$, $\sim 1.8 \times
10^{40}$ and $\sim 3.1 \times 10^{40}$ \es, respectively.  The soft
(hard) X-ray component provides about 30\% (70\%) of the total flux in
the soft energy band.  Almost all ($> 98\%$) the flux in the 2 - 10 keV
energy band is produced by the hard spectral component.  Given the
uncertainties on normalizations, the fluxes and luminosities reported
in Table 2 are estimated to be correct to about 30\%.


\subsection{The Soft X-ray Emitting Gas}

\subsubsection{Physical Parameters}

The straightforward interpretation of the origin of the soft X-ray spectral
component is to associate it with the diffuse X-ray emitting gas
that has been detected with the ROSAT PSPC and HRI data by A95.

The best fit spectral parameters reported in Table 1 allow us to
determine the basic physical conditions of this X-ray emitting gas.
For a gas with a temperature of $\sim 0.8$ keV and a normalization of
$\sim 1\times 10^{-4}$  at 1 keV \footnote { This number is equal to
$[10^{-14}/(4 \pi D^2)] \int n_e^2 dV$, where D is the distance to the
source in cm, $n_e$ is the electron density in units of cm$^{-3}$ and V
is the volume filled by the X-ray emitting gas in cm$^{3}$.} (a mean
value between the normalizations reported in Table 2), the emission
integral (EI = $\int n_e^2 dV$) is equal to $1.61 \times 10^{62}$
cm$^{-3}$.

For consistency we will adopt the same geometry as A95:
namely, half of the emission integral is represented by gas interior to
a radius of $\sim 0.5^{\prime} \simeq 1.69$ kpc (the ``inner region")
and the remaining is represented by the gas located between radii of $\sim
0.5^{\prime}$ and $\sim 2.5^{\prime}$, or 1.69 - 8.43 kpc (the
``outer region").

Parameterizing the clumpiness of the gas by a volume (V) filling factor
f, we can  determine:
the gas density 
($n_e \sim ({E.I. \over V \ f})^{1/2}$),
the gas pressure
($p \sim 2\ n_e \ k \ T$),
the gas mass
($M \sim  n_e \ m_p \ V \ f$),
the thermal energy of the gas
($E \sim 3 \ n_e \ k \ T \ V$) 
and the radiative cooling time 
($t_{cool} \sim {3 \ k \ T \over \wedge n_e}$), 
where T is the temperature of the gas and $\wedge$ is
the emissivity of the gas.  In computing the radiative cooling
time we have taken $\wedge = 2.0 \times 10^{-23}$ erg cm$^{3}$ s$^{-1}$
which is appropriate for gas in collisional ionization equilibrium
with solar abundances  and a
temperature of 0.8 keV (Sutherland and Dopita 1993).
The physical parameters of the soft X-ray emission implied by 
the best fit spectral model, in the ``inner" and the ``outer" regions, 
are reported in Table 3. 

We note that an increase of a factor 2 of the spectral normalization
implies an increase of a factor 1.41 in the physical parameters of the
X-ray emission reported in Table 3 (except for the radiative cooling
time, which decreases by the same factor).  Therefore the quantities
reported in Table 3 can be considered a good order-of-magnitude
estimate of the physical conditions of the diffuse soft X-ray emitting
gas in NGC 2146.

\subsubsection{The Origin of the Soft X-ray Emission}

We can now compare the above quantities to a simple model to check the
assertion of A95 that the soft, spatially-resolved X-ray emission in
NGC 2146 is produced by an outflowing superwind. To do so, we will
adopt a simple model that assumes that mechanical energy is being
continuously injected inside the starburst and used to heat the
surrounding gas. This gas then flows out of the starburst and emits
X-rays (cf. Wang 1995). In such a model we need to specify the heating
rate and the mass-injection rate into the outflow.
 
We begin by using starburst models of Leitherer and Heckman (1995 -
hereafter LH95) to scale the observed bolometric luminosity of NGC 2146
to obtain a predicted rate of mechanical energy injection (heating).
The total IR luminosity of the system is about $4 \times 10^{10}$
L$_{\odot}$.  The LH95 models then predict a total mechanical luminosity
of about $1.3 \times 10^{42}$ erg s$^{-1}$, and a star-formation rate
of $\sim$ 5 M$_{\odot}$ per year (for a Salpeter IMF extending from 0.1
to 100 M$_{\odot}$).
 
We will assume that the gas in the starburst is initially heated to a
temperature of about 10$^7$ K (the temperature of the soft component).
For the energy injection rate given above, conservation of energy then
implies that gas is heated at a rate of 9 M$_{\odot}$ per year. This is
about an order of magnitude greater than the predicted rate at which
supernovae and stellar winds return mass to the interstellar medium in
this system (cf. LH95). This implies that the outflow is strongly
``mass-loaded'': most of the outflowing material is ambient
interstellar gas in and around the starburst that has been heated by
the supernovae and stellar winds (cf. Suchkov et al 1996).  In this
case, we would expect the metallicity of the X-ray emitting gas to be
similar to that in the interstellar medium of NGC 2146 (see Ptak et
al., 1997 and Della Ceca, Griffiths and Heckman, 1997 for similar
results on NGC253, M82 and NGC 4449).
 
Since the energy injection rate of the starburst is several hundred
times the X-ray luminosity of the soft component, we will ignore the
dynamical effect of radiative cooling in our simple model. This neglect
is further justified since the estimated radiative cooling time of the
X-ray gas (of order 10$^9$ years - see Table 3) is much longer than the
adiabatic expansion timescale (of order 10$^7$ years).
 
Following Chevalier and Clegg (1985) and Wang (1995) we therefore
consider a spherically-symmetric wind with a mass outflow rate of 9
M$_{\odot}$ per year which is ``fed'' by hot gas that is injected at a
uniform temperature of 10$^7$ K.  We will take the size of the energy
injection region to be similar to that of the central radio source and
the bright X-ray emission in the ROSAT HRI data in A95 (radius $\sim$ 1
kpc).  The wind reaches a terminal velocity of v$_{wind}$ = 660 km
s$^{-1}$ (Chevalier and Clegg 1985) and cools through adiabatic
expansion while also emitting soft X-rays.  The emissivity of gas in
the 0.1 to 2.2 keV energy band is constant to within about a factor of
two as a function of temperature over the range between $8 \times 10^5$
K to 10$^{7}$ K ($\simeq 2 \times 10^{-23}$ erg cm$^{3}$ s$^{-1}$ for
solar metallicity).  This then allows us to predict an X-ray luminosity
from the outflow of $\simeq 1 \times 10^{40}$ erg s$^{-1}$ from within
a region equal in size to the X-ray nebula (radius of $\sim$7 Kpc),
plus an additional contribution of $\simeq 5 \times 10^{39}$ erg s$^{-1}$
from the hot gas inside the starburst that feeds the outflow.
The total estimated luminosity is about a factor of  three larger than
the measured luminosity of the soft X-ray component, which we regard as
satisfactory agreement for such a simple model.  Within the X-ray
nebula, the predicted mass of the hot gas will be about 10$^{8}$
M$_{\odot}$ and the predicted thermal plus kinetic energy in the
outflow will be $5 \times 10^{56}$ ergs. Again, these predicted values
agree reasonably well with the derived values for the sum of the inner
and outer regions (see Table 3): M $\simeq 7 \times 10^7 f^{1/2}$
M$_{\odot}$ and E$_{therm} \simeq 3 \times 10^{56} f^{1/2}$ ergs.

Note that a radius of 7 kpc and a wind terminal velocity of 660 km
s$^{-1}$ corresponds to a dynamical age of order 10$^7$ years
(reasonable for a starburst). If the outflow has lasted much longer
than this, the total mass and energy in the flow will rise accordingly,
but this will comprise material at larger radii ($>$ 7 kpc). Such
material will have a low density and (due to adiabatic cooling) a low
temperature, and would therefore not produce significant X-ray emission
(cf. Wang 1995).

\subsubsection{The Fate of the Hot Gas}
 
Having demonstrated that a very simple model can reproduce the gross
properties of the soft X-ray emission, we next consider the fate
of this outflowing gas. Following Wang et al. (1995), the ``escape
temperature'' for hot gas in a galaxy potential with an escape
velocity v$_{es}$ is given by:
 
T$_{es}$ $\simeq$ $1.26\times 10^{6}$ (v$_{es}$/300 km s$^{-1}$)$^{2}$ $ ^o$K
 
The HI rotation curve of NGC 2146 has an amplitude of 272 km s$^{-1}$.
Assuming that NGC 2146 has an isothermal dark matter halo extending to
a radius of 100 kpc, the escape velocity (see equation 13 in Heckman et
al. (1995)) at r = 7 kpc is about 740 km s$^{-1}$ and T$_{es}$ = $8
\times 10^6$ K. If the dark matter halo extends out to 300 kpc,
T$_{es}$ rises to $\sim 10^7$ K. These temperatures are similar to
the temperature we measure for the hot gas ($\simeq 10^7$
K).  The escape velocity is also similar to the estimated terminal
velocity for the wind (see above). These rough arguments suggest that
the outflowing gas {\it may} be able to escape the gravitational
potential of NGC 2146, and thereby carry metals and kinetic energy out
into the inter-galactic medium.

\subsection{The Origin of the Hard X-ray Emission} 

A95 discusses the following  scenarios for the origin of the hard 
spectral component:
a) thermal emission from the hot wind fluid that is built up 
   in the starburst injection region before the wind expands;
b) emission from a population of massive X-ray binaries and/or 
   young supernova remnants in the starburst;
c) emission from normal sources not associated with the 
   starburst;
d) Inverse Compton emission from the galactic wind and 
e) an optically obscured active galactic nucleus. 

They reach the conclusion that the two most plausible possibilities are
Inverse Compton Emission from the galactic wind and emission from a
population of massive X-ray binaries and/or young supernova remnants
associated with the starburst.  We will now discuss these two
possibilities.

Low resolution VLA observations of NGC 2146 show a diffuse radio halo
surrounding the central starburst; this radio halo is probably produced
by synchrotron emission from relativistic electrons that are flowing
out of the starburst.  Since the starburst also produces a plentiful
supply of ``soft" (ultraviolet, optical and infrared) photons, inverse
Compton scattering of these soft photons by the relativistic electrons
will lead to non-thermal X-ray emission from the halo.  Furthermore,
the hard X-ray spectral component can be satisfactorily described by a
power law model (best fit photon index of $\sim$ 1.7), which is highly
suggestive since  this is the X-ray spectral signature of the
Compton-scattered photons.  It is therefore important to make a rough
estimate of the expected inverse Compton luminosity of NGC 2146.

The inverse Compton luminosity ($L_{IC}$) 
and the Synchrotron luminosity ($L_{Syn}$)
are tied together by the relationship,  
$ L_{IC} = L_{Syn} \times {U_{rad} \over U_{B}}$, 
where  $U_{rad}$ is the energy  density of the radiation field and 
$U_{B} (= B^{2}/8\pi)$ is the energy density of the magnetic field (B).
$U_{B}$ can be estimated by applying a ``minimum" energy calculation 
(see Moffet, 1975).
Input parameters are:
a) $L_{Syn} = 1.53 \times 10^{39}$ \es.  This latter represents the
total radio luminosity (integrated over the frequency range from 10 MHz
to 10 GHz) obtained by using a 1.4 GHz flux of $\sim 1.3$ Jy (White et
al., 1992) and assuming a spectral index of 0.7;
b) a ratio of total energy of CRs to total energy of electrons k=100;
c) a volume emitting region approximated by a cylinder with a height of 
   $\sim $ 0.8 kpc and a radius of $\sim $ 1.4 kpc as estimated from the 
   high frequency and high resolution map reported in Lisenfeld et al.    
   (1996). 
With these  assumptions we obtain  
a magnetic field energy density 
$U_{B} =  1.1 \times 10^{-10}$ erg cm$^{-3}$ (B = 52.8 $\mu G$).
$U_{rad}$ can be computed from the bolometric luminosity, 
$U_{rad} = {L_{bol}\over 4 \pi R^{2} c}$.
Since most of the luminosity in NGC 2146 is emitted in the far 
infrared we will assume 
$L_{bol} \simeq L_{1-100 \mu m} \simeq 1.5 \times 10^{44}$ \es. 
For the radius R, we will take R=1.4 kpc, which represents the radius 
of the radio halo.
With these  assumptions we obtain 
$U_{rad} =  2.1 \times 10^{-11}$ erg cm$^{-3}$.
Combining these values we have 
$ L_{IC:2-10 keV} \lae 3 \times 10^{38}$ \es.
Therefore it is clear that, unless we are far from the minimum-energy
conditions in the radio source, the inverse Compton process will
produce only a few percent of the X-ray luminosity of the hard X-ray
component.
Note that if we use k=1 instead of k=100, then 
$U_{B} = 1.2 \times 10^{-11}$ erg cm$^{-3}$ (B = 17.2 $\mu G$)
and $L_{IC:2-10 keV} \lae 2.7 \times 10^{39}$ \es:
this is still not enough to produce the hard X-ray emission in NGC 2146.

Excluding inverse Compton emission we are then left with the emission
from a population of massive X-ray binaries and/or young supernova
remnants associated with the starburst.  The total (0.2 - 2.4) keV
X-ray flux of the four brightest X-ray sources resolved by the ROSAT
HRI in the central starburst of NGC 2146 is $\sim 3-4\times 10^{-13}$
\ecs.  The ASCA best fit model(s) imply an (0.2 - 2.4 keV)
X-ray flux of the hard component which is a factor two higher than this.
Given the uncertainties in the ROSAT HRI fluxes and considering the 
``first order" nature of the NGC 2146 X-ray spectra, we 
consider highly plausible that the hard X-ray component represents
the combined emission of the ``point-like" sources in the starburst. 
On the other hand, it is worth noting that while massive X-ray binaries
and/or young supernova remnants are plausible 
counterpart of these ``point-like" sources, 
their average spectral properties are different from a
power-law model with photon index equal to 1.7 or from a thermal
spectrum with kT$\sim$ 8 keV.  The massive X-ray binaries have spectra
too hard (photon index in the range 0.8-1.5 below 10 keV; Nagase,
1989), while the young supernova remnants are too soft (kT $\sim$ 2-3 keV).
The only class of discrete X-ray sources that could have a power-law
spectrum with photon index around 1.7 are black-hole binaries in the
low state (Tanaka, 1996)
\footnote{
It is worth noting that a power-law spectrum with photon index around
1.7 is also consistent with an AGN hypothesis. However this possibility
was discussed and ruled out by A95 on the basis of a) the size of the
X-ray nebulae in the 1-2 kev energy range and b) the starburst nuclear
spectrum of NGC 2146.}
. About two to five thousand systems emitting
at the low state luminosity (several times 10$^{36}$ \es) should be
required to take into account of the hard spectral component of NGC 2146.
Given that about 1000 (or even more) of such system are estimated 
to be present in our Galaxy (see Tanaka, 1996), such possibility can not 
be excluded for a starburst galaxy like NGC 2146.

In summary, while we are confident that the hard X-ray emission is due
to the combined emission of the point-like sources associated with the
starburst, to understand their physical nature we will have to wait for
JETX, AXAF and/or XMM observations.

\section{Summary and Conclusion}

In this paper we have presented and discussed ASCA 
observations of the starburst galaxy NGC 2146. 

The main results of this paper are summarized as follows:

a)  The broad band (0.6-9 keV) X-ray spectrum of NGC 2146 is well
fitted by a two component model: the soft component can be described by a
Raymond-Smith thermal model with a temperature of $\sim 0.8$ keV, while
the hard component can be described by a thermal model with a
temperature of $\sim 8$ keV or by a power-law model with photon index
of $\sim 1.7$. The $N_{H}$ along the line of sight is consistent with 
the Galactic value. 
Similar spectral components are found to describe the broad band 
X-ray spectral properties of other bright starburst galaxies 
(Ptak et al., 1997; Moran and Lehnert, 1997; Zezas, Georgantopoulos and 
Ward, 1998; Persic et al., 1998).

b) The total luminosity of NGC 2146 in the soft (0.5 - 2.0 keV), hard
(2 - 10 keV), and broad (0.5-10.0 keV) energy band is $\sim 1.3 \times
10^{40}$, $\sim 1.8 \times 10^{40}$ and $\sim 3.1 \times 10^{40}$ \es,
respectively.  The soft (hard) spectral component provides about 30\%
(70\%) of the total luminosity in the soft 0.5 - 2.0 keV energy band,
while in the hard 2-10 keV energy range only the hard spectral
component plays a major role.

c) We have shown that a starburst-driven ``mass-loaded" superwind model
(cf. Suchkov et al., 1996) can satisfactorily account for the gross
properties (luminosity, mass, and energy content) of the soft X-ray
spectral component. The estimated outflow velocity of the hot gas is
close to the escape velocity from the galaxy, so its fate is not
clear.

d) We have excluded Inverse Compton emission as
the dominant mechanism for the production of the hard spectral
component (unless we are far from the minimum-energy conditions in the
radio source).  We suggest that the hard spectral component is due to the
combined emission of massive X-ray binaries and/or young supernova
remnants associated with the starburst.

The new ASCA spectral data allowed us to confirm and extend  many of the
conclusions of A95 (which were based on ROSAT PSPC spectral and imaging
data and ROSAT HRI imaging data) in several important ways.
First, the broad band (0.6-9 keV) X-ray spectrum of NGC 2146
allows us to verify spectroscopically the presence of multiple
components in the X-ray emission, to measure their respective
luminosities and to evaluate their proportional contribution in the
soft ($E<2$ keV) and hard ($E> 2$ keV) energy bands.
Second, using the best fit spectral parameters we have been able to set
tighter constraints on the starburst-driven superwind model, in term of
mass outflow rate and outflow velocity.

Acknowledgments

The authors acknowledge partial financial support from NASA grants
NAG5-6400, NAG5-3651 and NAG5-6075.  We would like to thank
L. Armus and the referee, A. Ptak, for helpful comments and
suggestions.  This research has made use of the NASA/IPAC extragalactic
database (NED), which is operated by the Jet Propulsion Laboratory,
Caltech, under contract with the National Aeronautics and Space
Administration.  We thank all the members of the ASCA team who operate
the satellite and maintain the software data analysis and the archive.

%
%
%
%

\clearpage
{
\scriptsize
\begin{table*}
\begin{center}
\tablenum{1}
\caption{Results of the Spectral Fits Using Two Component Models.} 
\begin{tabular}{llccccc}
Model = Ray+Ray \\
Instruments  & $kT_{Soft}$           & Norm$_{Soft}$                  
             & $kT_{Hard}$           & Norm$_{Hard}$                                    
             & $N_{H}$               &$\chi^2_{\nu}$/(d.o.f.) \\
 (1)         & (2)                   & (3)                            
             & (4)                   & (5)                  
             & (6)                   & (7) \\
\tableline
G23+S01      & $0.82_{0.76}^{0.87}$  & $1.59_{1.11}^{2.05}$/$0.75_{0.49}^{0.97}$         
             & $8.34_{6.05}^{11.76}$ & $9.20_{8.44}^{9.99}$/$7.29_{6.83}^{7.81}$                  
             & 7.3(fixed)            & 1.12/252 \\
G23+S01+PSPC & $0.77_{0.66}^{0.86}$  & $1.85_{1.08}^{3.65}$/$0.88_{0.51}^{1.90}$/$1.65_{0.82}^{3.39}$        
             & $7.07_{5.15}^{11.2}$  & $9.41_{8.52}^{10.4}$/$7.53_{6.93}^{8.34}$/$6.28_{3.28}^{9.33}$  
             & $13_{2.5}^{41}     $  & 1.12/270 \\
\tableline
$\ $         & $\ $                  & $\ $                                             
             & $\ $                  & $\ $                            
             & $\ $                  & $\ $ \\
$\ $         & $\ $                  & $\ $                                             
             & $\ $                  & $\ $                            
             & $\ $                  & $\ $ \\
Model = Ray+Po   \\           
Instruments  & $kT_{Soft}$          & Norm$_{Soft}$ 
             & $\Gamma$             & Norm$_{Hard}$                                    
             & $N_{H}$              & $\chi^2_{\nu}$/(d.o.f.) \\
 (1)         & (2)                  & (3)                                         
             & (4)                  & (5)                                                             
             & (6)                  & (7) \\
\tableline
G23+S01      & $0.82_{0.70}^{0.89}$ & $1.29_{0.82}^{1.79}$/$0.57_{0.32}^{0.80}$         
             & $1.68_{1.57}^{1.79}$ & $3.00_{2.50}^{3.40}$/$2.31_{2.05}^{2.58}$                       
             & 7.3(fixed)           & 1.09/252 \\
G23+S01+PSPC & $0.73_{0.62}^{0.83}$ & $1.96_{1.08}^{7.00}$/$0.95_{0.45}^{4.20}$/$1.86_{0.93}^{6.97}$        
             & $1.79_{1.64}^{1.94}$ & $3.43_{2.95}^{4.09}$/$2.67_{2.25}^{3.21}$/$2.11_{1.17}^{3.23}$  
             & $21_{10}^{43}      $ & 1.06/270  \\
\tableline
\end{tabular}
\end{center}



%

\tablecomments{
\scriptsize
The elemental abundances in these fits have been fixed at the Solar values.
Allowed ranges are at $90\%$ confidence intervals for two interesting parameters
($\chi^2_{min} + 4.61$).  
Columns are as follows:
(1) Instrument; 
(2) Best fit and $90\%$ confidence interval for the temperature of the soft 
component in keV; 
(3)  Best fit and $90\%$ confidence interval for the normalization of the 
soft component at 1 keV in units of $10^{-4}$ for the G23, S01 and PSPC data set respectively. 
This number is equal to $[10^{-14}/(4 \pi D^2)] \int n_e^2 dV$, where D is the 
distance to the source in cm, $n_e$ is the electron density in units of cm$^{-3}$ and 
V is the volume filled by the X-ray emitting gas in cm$^{3}$.
(4) Best fit and $90\%$ confidence interval for  the temperature of the hard 
component (Ray+Ray model) in keV  or for the power law photon index (Ray+Po model); 
(5)  Best fit and $90\%$ confidence interval for the  normalization of the 
hard component at 1 keV in units of $10^{-4}$ (Ray+Ray model) 
or for the normalization of the 
power law component in units of $10^{-4}$ photons s$^{-1}$ cm$^{-2}$ keV$^{-1}$ at 1 keV (Ray+Po model).
G23, S01 and PSPC data set, respectively.
(6) The absorbing column density along the line of sight in units of 
$10^{20}$ cm$^{-2}$.
(7)  Reduced chi-squared and degree of freedom.
}
\end{table*}
}

{
\scriptsize
\begin{table*}
\begin{center}
\tablenum{2}
\caption{Unabsorbed X-ray Fluxes and luminosities} 
\begin{tabular}{lcccccc}
               Model      &  $kT_{Soft}$            & Norm$_{Soft}$  & $F_{x}$ ($L_{x}$)/$F_{x}$ ($L_{x}$) 
                          &  $kT_{Hard}$/$\Gamma$   & Norm$_{Hard}$  & $F_{x}$ ($L_{x}$)/$F_{x}$ ($L_{x}$)     \\
                          &                         &                &  (0.5 - 2.0) kev / (2.0 - 10.0) keV  
                          &                         &                &  (0.5 - 2.0) keV / (2.0 - 10.0) keV     \\
               (1)        & (2)                     & (3)            & (4)                
                          & (5)                     & (6)            & (7)                                     \\
\tableline
               Ray+Ray    & 0.82                    & 1.17           & 2.79 (4.49) / 0.16 (0.26)                      
                          & 8.34                    & 8.25           & 4.97 (8.00) / 11.38 (18.32)              \\
               Ray+Po     & 0.82                    & 0.93           & 2.22 (3.57) / 0.13 (0.20)                      
                          & 1.68                    & 2.66           & 5.96 (9.60) / 11.20 (18.03)              \\
\tableline
\end{tabular}
\end{center}



%

\tablecomments{
\scriptsize
Columns are as follows:
(1) Model; 
(2) Temperature of the soft component in keV ; 
(3)  Normalization of the soft component at 1 keV in units of $10^{-4}$. 
     This number is equal to $[10^{-14}/(4 \pi D^2)] \int n_e^2 dV$, where D is the 
     distance to the source in cm, $n_e$ is the electron density in units of cm$^{-3}$ and 
     V is the volume filled by the X-ray emitting gas in cm$^{3}$;
(4)  Fluxes and luminosities of the soft spectral component in the (0.5-2.0) keV energy band and in the 
     (2.0 - 10.0) keV energy band. The fluxes are in units of $10\time 10^{-13}$ \ecs, while 
     the luminosities are in units of $10\time 10^{39}$ \es;
(5)  Temperature of the hard component (Ray+Ray model) in keV or the power law photon index (Ray+Po model);
(6)  Normalization of the hard component at 1 keV in units of $10^{-4}$ (Ray+Ray model) 
     or in units of $10^{-4}$ photons s$^{-1}$ cm$^{-2}$ keV$^{-1}$ at 1 keV (Ray+Po model);
(7)  Fluxes and luminosities of the hard spectral component in the (0.5-2.0) keV energy band and in the 
     (2.0 - 10.0) keV energy band. The fluxes are in units of $10\time 10^{-13}$ \ecs, while 
     the luminosities are in units of $10\time 10^{39}$ \es.
Numbers reported in column 2 and 5 are the best fit parameters of the
combined ASCA data set (with the $N_{H}$ held fixed at the Galactic value)
for the soft and hard component. respectively.
Their corresponding normalizations (column 3 and 6) are mean values
between the G23 and S01 normalizations (see Table 1).
}
\end{table*}
}

\clearpage
{
\scriptsize
\begin{table*}
\begin{center}
\tablenum{3}
\caption{Physical Conditions and Energetics of the Soft X-ray Emitting Gas} 
\begin{tabular}{lcccccc}
Region     & Volume              & $<n_e>$                        &  $P_x$                        & $M_x$    
           & $E_x$                                                & $t_{cool}$                              \\
           & cm$^{3}$            & cm$^{-3}$                      & dyne cm$^{-2}$                & M$_{\odot}$
           & erg                                                  & years                                    \\
(1)        & (2)                 & (3)                            &  (4)                          & (5) 
           & (6)                 & (7)                                                                       \\
\tableline
inner      & $5.9\times 10^{65}$ & $1.2\times 10^{-2} f^{-1/2}$   & $3.0\times 10^{-11} f^{-1/2}$ & $6.8\times 10^{6} f^{1/2}$ 
           & $2.6\times 10^{55} f^{1/2}$                          & $5.2\times 10^{8} f^{1/2}$                \\
outer      & $7.3\times 10^{67}$ & $1.0\times 10^{-3} f^{-1/2}$   & $2.7\times 10^{-12} f^{-1/2}$ & $6.4\times 10^{7} f^{1/2}$ 
           & $2.9\times 10^{56} f^{1/2}$                          & $5.8\times 10^{9} f^{1/2}$                \\
\tableline
\end{tabular}
\end{center}



%

\end{table*}
}

%

\clearpage
\section{Figure Captions}

Figure 1: Residuals to the best fit Raymond-Smith model. 
G23: open squares; S01: filled squares.  For this plot, the
spectra have been rebinned so that the signal-to-noise ratio in each
channel is at least 3.

Figure 2: The folded spectrum  and the residuals of the 
GIS2+GIS3 data set compared with the best fit Raymond-Smith plus 
Power Law model. For this plot, the spectra have been rebinned 
so that the signal-to-noise ratio in each channel is at least 3.

Figure 3: The folded spectrum  and the residuals of the 
SIS0+SIS1 data set compared with the best fit Raymond-Smith plus 
Power Law model. For this plot, the spectra have been rebinned 
so that the signal-to-noise ratio in each channel is at least 3.

Figure 4: The unfolded spectrum of the ASCA + ROSAT PSPC data set
(Ray+Po model).  For this plot, the spectra have been rebinned so that
the signal-to-noise ratio in each channel is at least 3.

\clearpage
\begin{figure}
\figurenum{1}
\plotone{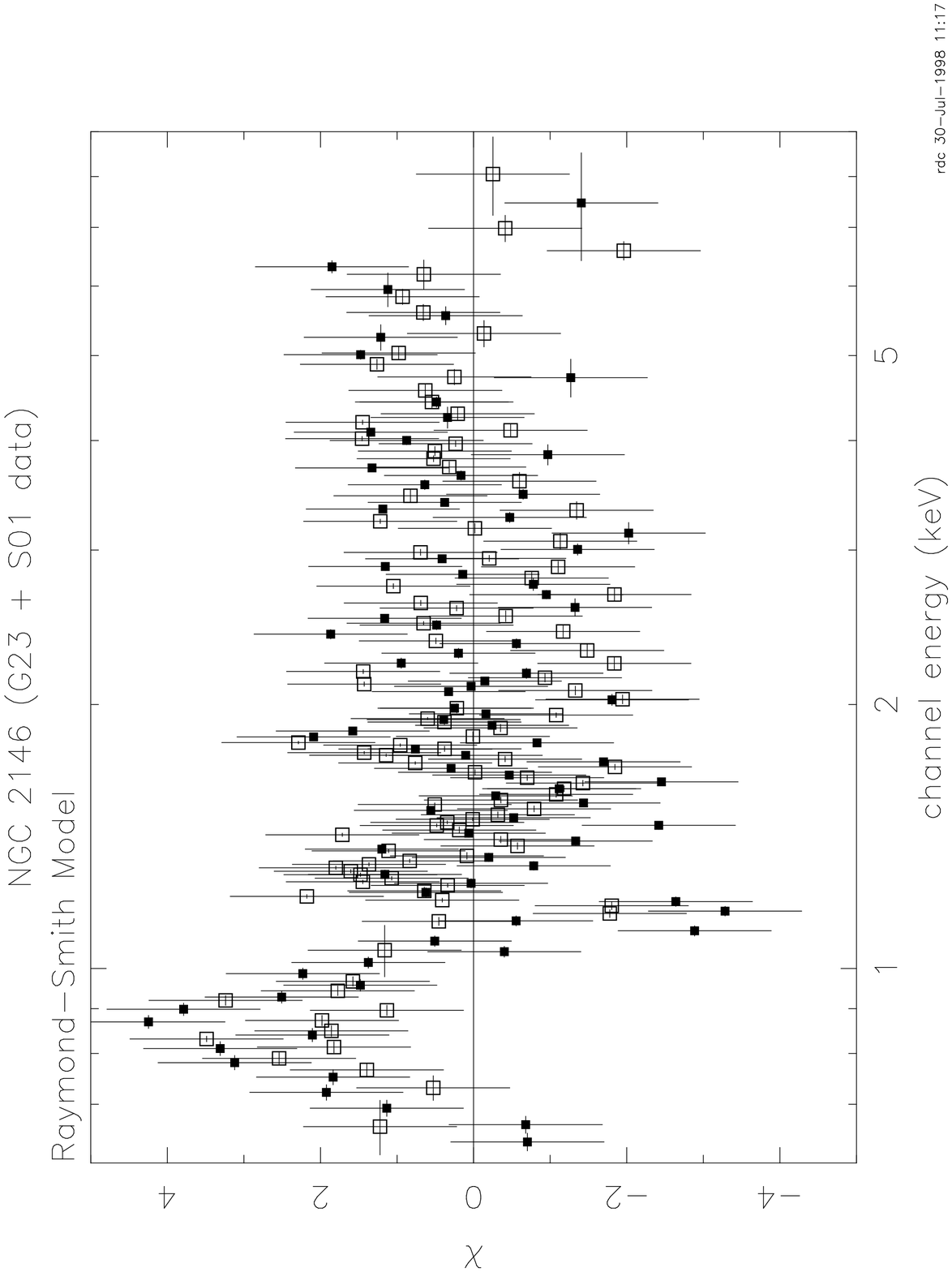}
\caption{}
\end{figure}

\clearpage
\begin{figure}
\figurenum{2}
\plotone{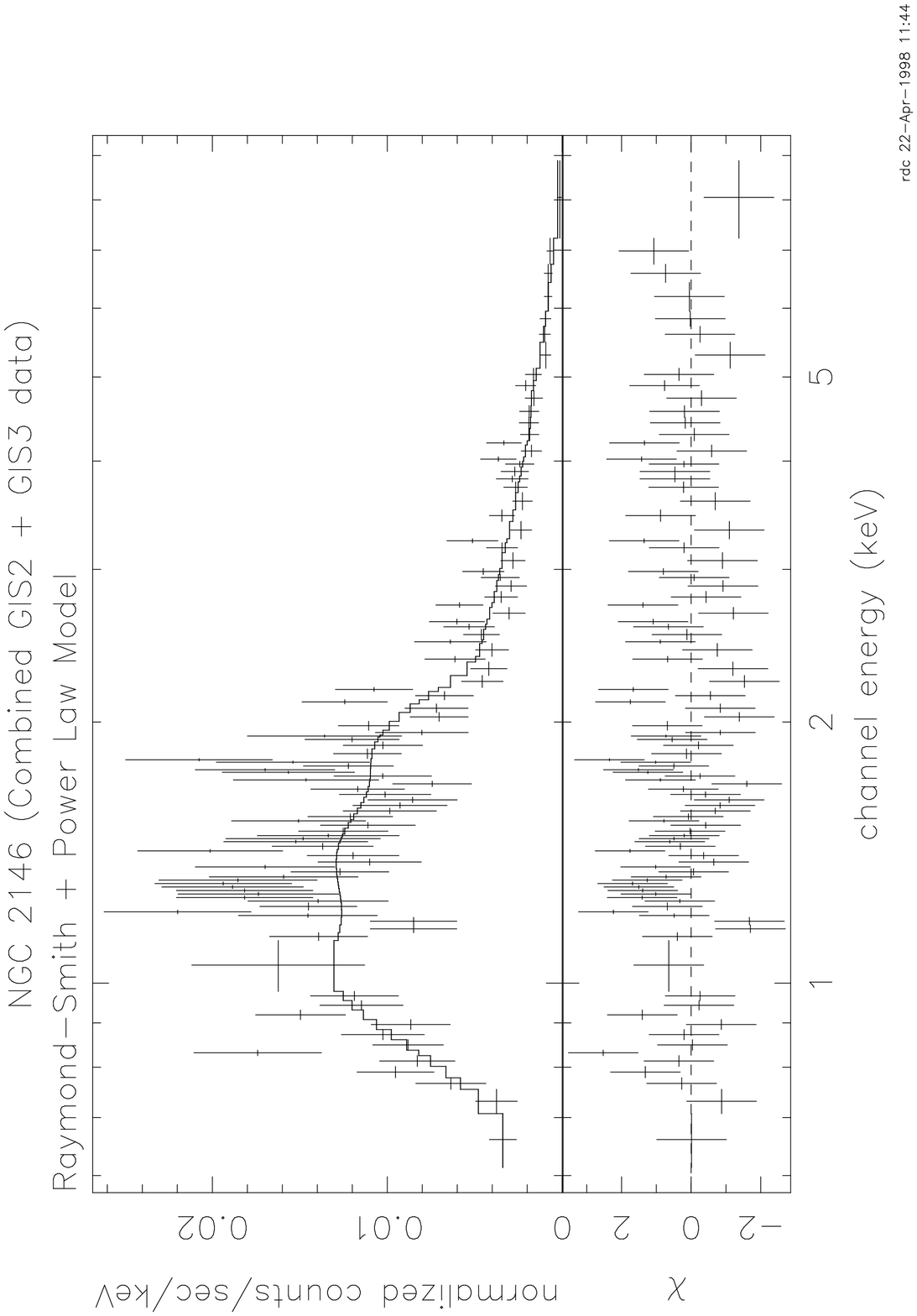}
\caption{}
\end{figure}

\clearpage
\begin{figure}
\figurenum{3}
\plotone{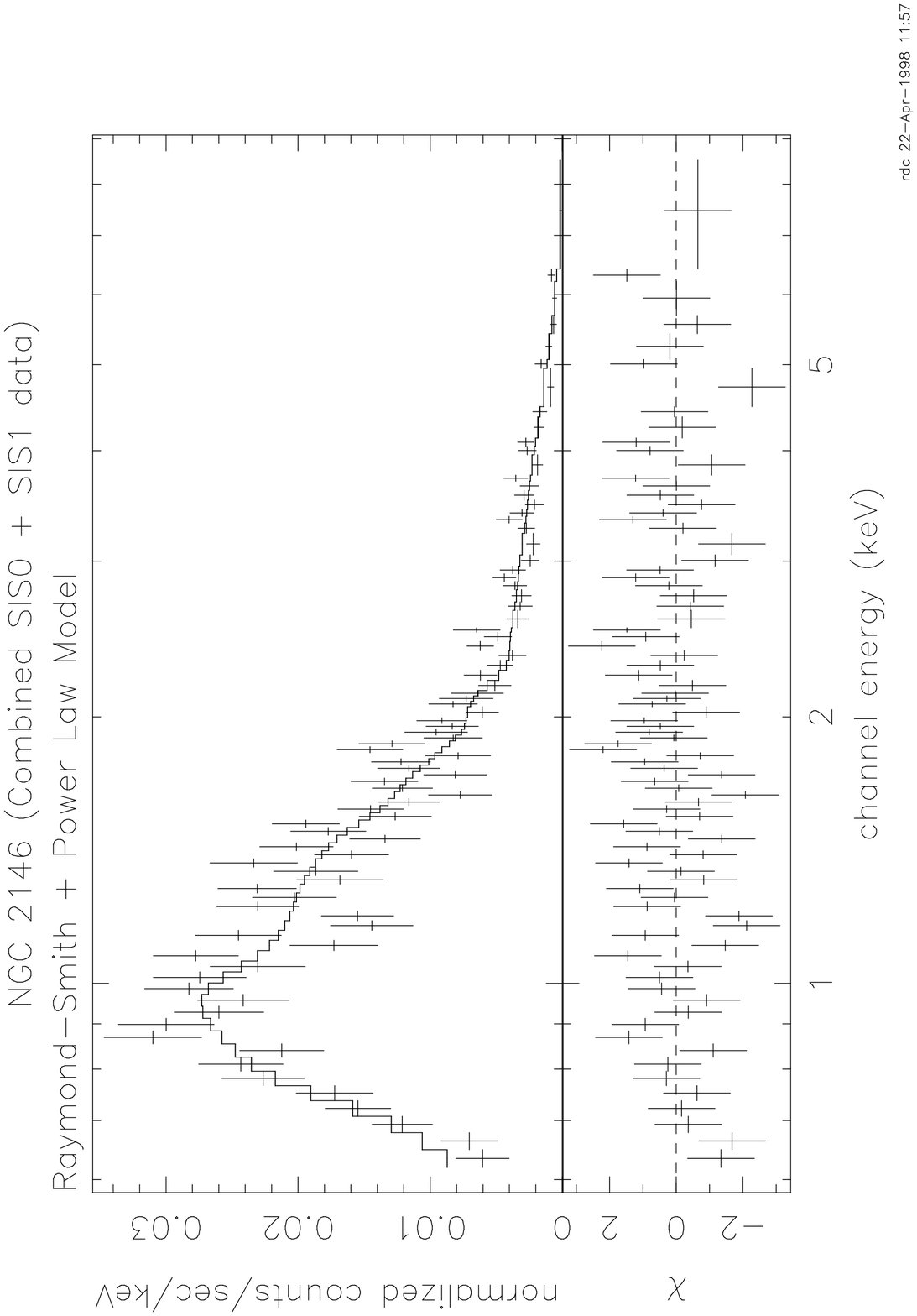}
\caption{}
\end{figure}

\clearpage
\begin{figure}
\figurenum{4}
\plotone{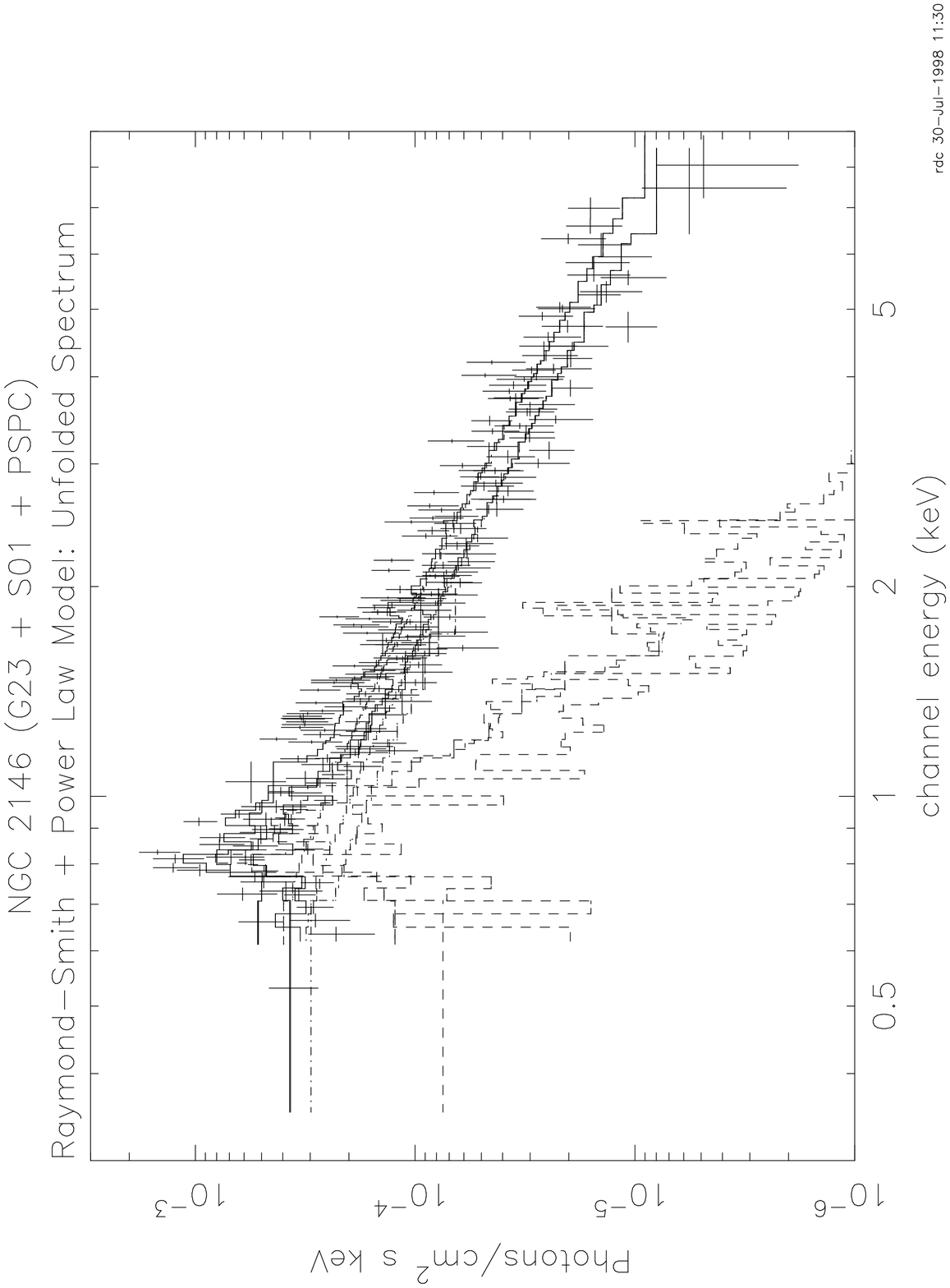}
\caption{}
\end{figure}

%


\clearpage 

\begin{thebibliography}{}


\bibitem{} \reference Armus, L., Heckman, T.M., Weaver, K.A., and Lehnert,
 M.D. 1995, \apj, 445, 666.

\bibitem{} \reference Arnaud, K.A. 1996, ASP Conf.Series, 101,17.










\bibitem{} \reference Bland-Hawthorn, J., Publ.Astron.Soc.Aust., 12, 190.














\bibitem{} \reference Della Ceca, R., Griffiths, R.E., Heckman, T.M., and 
MacKenty, J.W. 1996, \apj, 469, 662.

\bibitem{} \reference Della Ceca, R., Griffiths, R.E., Heckman, T.M., 1997, \apj, 485, 581.



\bibitem{} \reference de Vaucouleurs, G., de Vaucouleurs, A., Corwin,
H. Buta, R., Paturel, R., and Fouque, P. 1991, Third Reference Catalog
of Bright Galaxies (Austin: Univ. Texas Press).

\bibitem{} \reference Dahlem, M., Weaver, K.A., and Heckman, T.M. 1998, 
\apjs, in press. 


\bibitem{} \reference Dotani, T., et al., 1996, ASCA News, 4, 3.








\bibitem{} \reference George, I.M., Arnaud, K.A., Pence, B., and
Ruamsuwan, L. 1992, Legacy, 2, 51.









\bibitem{} \reference Heckman, T.M., Lehnert, M.D., and Armus, L. 1993, 
in The Evolution of Galaxies and Their Environments, ed. S.M. Shull and 
H. Thronson (Dordrecht: Kluwer), 455.
















\bibitem{} \reference Lehnert, M.D., and Heckman, T.M. 1995, \apjs, 97, 89.

\bibitem{} \reference Lehnert, M.D., and Heckman, T.M. 1996, \apj, 462, 651.









\bibitem{} \reference Lisenfeld, U., Alexander, P., Pooley, G.G., and Wilding, 
T. 1996, \mnras, 281, 301.











\bibitem{} \reference Moffet, A.T. 1975 in Stars and Stellar System, vol. IX:
Galaxies and the Universe, eds A.Sandage, M.Sandage and J.Kristian 
(University of Chicago Press:Chicago),p. 211.

\bibitem{} \reference Moran, E.D., and Lehnert, M.D. 1997, \apj, 478, 172.

\bibitem{} \reference Nagase, F., 1989, \pasj, 41, 1.






\bibitem{} \reference Persic, M., et al., 1998, astro-ph/9809256.

\bibitem{} \reference Ptak, A., Serlemitsos, P., Yaqoob, T., Mushotzky, R., 
and Tsuru, T. 1997, \aj, 113, 1286.









 



\bibitem{} \reference Suchkov, A.A., Balsara, D.S.,  Heckman, T.M., and 
Leitherer, C. 1994, \apj, 430, 511.




\bibitem{} \reference Tanaka, Y., Inoue, H., and Holt, S.S. 1994, PASJ, 46, L37.

\bibitem{} \reference Tanaka, Y., 1996, MPE Report, 263, 85.







\bibitem{} \reference White, R.L., et al. 1992, \apjs, 79, 331.






\bibitem{} \reference Zezas, A.L., Georgantopoulos, I., and Ward, M.J. 1998,
astro-ph/9807258

\end{thebibliography}
\end{document}